\documentclass[aps,prl,showpacs,twocolumn,superscriptaddress]{revtex4}
\usepackage{bbm}
\usepackage{mathrsfs}
\usepackage{epsfig,psfrag}
\usepackage{amsmath,amsfonts,amssymb}
\usepackage[usenames]{color}
\usepackage{bm}

\newcommand{\etal}[1]{{#1}}

\begin{document}

\title{Casimir Force at a Knife's Edge}

\author{Noah Graham}
\email{ngraham@middlebury.edu}
\affiliation{Department of Physics,
Middlebury College,
Middlebury, VT 05753, USA}

\author{Alexander Shpunt}
\affiliation{Department of Physics, Massachusetts Institute of
Technology, Cambridge, MA 02139, USA}

\author{Thorsten Emig}
\affiliation{Department of Physics,
Massachusetts Institute of Technology,
Cambridge, MA 02139, USA}
\affiliation{Institut f\"ur Theoretische Physik,
Universit\"at zu K\"oln,
Z\"ulpicher Strasse 77,
50937 K\"oln, Germany}
\affiliation{Laboratoire de Physique Th\'eorique et Mod\`eles Statistiques,
CNRS UMR 8626, B\^at.~100,
Universit\'e Paris-Sud,
91405 Orsay cedex,
France}

\author{Sahand Jamal Rahi}
\affiliation{Department of Physics, Massachusetts Institute of
Technology, Cambridge, MA 02139, USA}

\author{Robert L. Jaffe}
\affiliation{Department of Physics, Massachusetts Institute of
Technology, Cambridge, MA 02139, USA}
\affiliation{Center for Theoretical Physics and Laboratory for Nuclear
Science, Massachusetts Institute of
Technology, Cambridge, MA 02139, USA}

\author{Mehran Kardar}
\affiliation{Department of Physics, Massachusetts Institute of
Technology, Cambridge, MA 02139, USA}

\begin{abstract}
The Casimir force has been computed exactly for only a few simple
geometries, such as infinite plates, cylinders, and spheres.  We show
that a parabolic cylinder,
for which analytic solutions to the Helmholtz equation are available,
is another case where such a calculation
is possible.  We compute the interaction energy of a parabolic cylinder
and an infinite plate (both perfect mirrors), as a function of
their separation and inclination, $H$ and $\theta$, and the cylinder's
parabolic radius $R$.  As $H/R\to 0$, the proximity force
approximation becomes exact. The opposite limit of $R/H\to 0$
corresponds to a semi-infinite plate, where the effects of edge
and inclination can be probed.

\end{abstract}

\pacs{42.25.Fx, 03.70.+k, 12.20.-m}

\maketitle

Casimir's computation of the force between two parallel metallic
plates~\cite{Casimir48} originally inspired much theoretical interest
as a macroscopic manifestation of quantum fluctuations of the
electromagnetic field in vacuum.  Following its experimental
confirmation in the past decade~\cite{experiments}, however, it is now an
important force to reckon with in the design of microelectromechanical
systems~\cite{MEMS}.  Potential practical applications have motivated
the development of numerical methods to compute Casimir forces for
objects of any shape~\cite{Johnson}.  The simplest and most commonly
used methods for dealing with complex shapes rely on pairwise
summations, as in the proximity force approximation (PFA), which
limits their applicability.

Recently we have developed a formalism~\cite{spheres,universal} 
that relates the Casimir interaction among several objects to the
scattering of the electromagnetic field from the objects individually.
(For additional perspectives on the scattering formalism, see
references in~\cite{universal}.)  This approach simplifies the
problem, since scattering is a well-developed subject.  In
particular, the availability of scattering formulae for simple
objects, such as spheres and cylinders, has enabled us to  compute the
Casimir force between two spheres~\cite{spheres}, a sphere and a
plate~\cite{sphere+plate}, multiple cylinders~\cite{cylinders}, etc.
In this work we show that \emph{parabolic} cylinders provide another
example where the scattering amplitudes can be computed exactly.  We
then use the exact results for scattering from perfect mirrors to
compute the Casimir force between a parabolic cylinder and a plate.
In the limiting case when the radius of curvature at its tip vanishes,
the parabolic cylinder becomes a semi-infinite plate (a knife's edge),
and we can consider how the energy depends on the boundary condition
it imposes and the angle it makes to the plane.

\begin{figure}[htbp]
\includegraphics[width=0.75\linewidth]{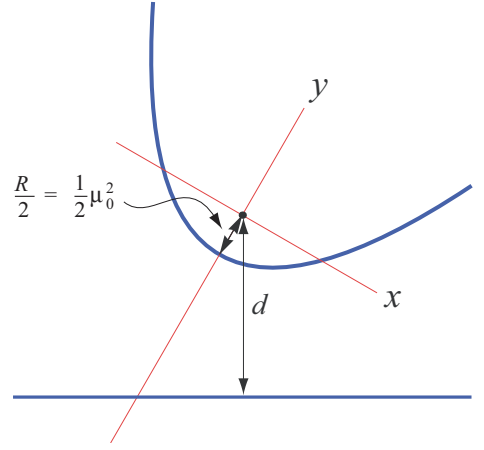}
\caption{Parabolic cylinder/plane geometry.}
\label{fig:tilt}
\end{figure}

The surface of a parabolic cylinder in Cartesian coordinates is
described by $y=(x^2-R^2)/2R$ for all $z$, as shown in
Fig.~\ref{fig:tilt}, where $R$ is the radius of curvature at the tip.
In parabolic cylinder coordinates~\cite{MF}, defined
through $x=\mu\lambda$, $y=(\lambda^2-\mu^2)/2$, $z=z$, the surface is
simply $\mu=\mu_0=\sqrt{R}$ for  $-\infty<\lambda,z<\infty$.  One
advantage of the latter coordinate system is that the Helmholtz
equation
\begin{equation}
\nabla^2 \Phi = \frac{1}{\lambda^2 + \mu^2} 
\left(\frac{d^2 \Phi}{d\lambda^2} + \frac{d^2 \Phi}{d\mu^2} \right) 
+ \frac{d^2 \Phi}{dz^2} =\kappa^2 \Phi \,,
\end{equation}
which we consider for imaginary wavenumber $k=i\kappa$,
admits separable solutions.  Since sending $\lambda \to -\lambda$ and
$\mu \to -\mu$ returns us to the same point, we restrict our attention
to $\mu\ge 0$ while considering all values of $\lambda$.  Then $\mu$
plays the role of the ``radial'' coordinate in scattering theory and
we have regular and outgoing wave solutions
\begin{eqnarray}
\psi_\nu^{\hbox{\tiny reg}}(\bm{r}) &=& 
i^{\nu} e^{i k_z z} D_\nu(\tilde \lambda)  D_\nu(i\tilde \mu) \,, \cr
\psi_\nu^{\hbox{\tiny out}}(\bm{r}) &=& 
e^{i k_z z} D_\nu(\tilde \lambda)  D_{-\nu-1}(\tilde \mu) \,,
\end{eqnarray}
where $D_\nu(u)$ is the parabolic cylinder function, and
$\tilde \lambda= \lambda \sqrt{2\sqrt{k_z^2 + \kappa^2}}$ and
similarly for $\mu$.  Enforcing the reflection symmetry $\lambda
\to -\lambda$ and $\mu \to -\mu$ for the regular solutions restricts
the separation constant $\nu$ to integer values.  The corresponding
outgoing solutions do not obey this restriction and thus can only be
used away from $\mu=0$; as is typical for outgoing solutions, they are
irregular at $\mu=0$.  For imaginary wavenumber, the regular (outgoing)
solutions grow (decay) exponentially in $\mu$ and both
$i^\nu D_\nu(i\tilde \mu)$ and $D_\nu(\tilde \lambda)$
are real.  We can then express the free scalar Green's 
function as~\cite{MF}
\begin{equation}
G(\bm{r}_1, \bm{r}_2, \kappa)
= \int_{-\infty}^\infty \frac{d k_z}{2 \pi}
\sum_{\nu=0}^\infty \frac{(-1)^\nu}{\nu!\sqrt{2\pi}}
\psi_\nu^{\hbox{\tiny reg}}(\bm{r}_<)^\ast
\psi_\nu^{\hbox{\tiny out}}(\bm{r}_>) \,,
\label{eqn:Green1}
\end{equation}
where $\bm{r}_<$ ($\bm{r}_>$) is the coordinate with the smaller
(larger) value of $\mu$.  We will also use the Green's function in
coordinates appropriate to scattering from a plane perpendicular to
the $y$-axis,
\begin{eqnarray}
G(\bm{r}_1, \bm{r}_2, \kappa) &=& 
\int_{-\infty}^\infty \frac{d k_z}{2 \pi} e^{i k_z(z_2 - z_1)} \cr
&\times& \frac{i}{4\pi} \int_{-\infty}^\infty \frac{dk_x}{k_y} 
e^{ik_x (x_2-x_1) + i k_y |y_2 - y_1|} \,,
\label{eqn:Green2}
\end{eqnarray}
where $k_y = i\sqrt{\kappa^2+k_x^2+k_z^2}$.  We can connect the 
parabolic and Cartesian Green's functions using the expansion of a
plane wave in regular parabolic solutions~\cite{MF}
\begin{equation}
e^{i\bm{k}\cdot\bm{r}} =  
\sum_{\nu=0}^\infty \frac{1}{\nu!}
\frac{\left(\tan \frac{\phi}{2} \right)^{\nu}}
{\cos \frac{\phi}{2}}
\psi_\nu^{\hbox{\tiny reg}}(\bm{r}),
\label{eqn:plane1}
\end{equation}
where $\tan \phi = \frac{k_x}{k_y}$.  This expression converges in
regions where $|\tan \frac{\phi}{2}| < 1$.  A plane wave with $|\tan
\frac{\phi}{2}| > 1$ can instead be expanded in terms of solutions with 
negative integer values of $\nu$~\cite{MF}, and the Green's function
can also be expressed in terms of these functions analogously to
Eq.~(\ref{eqn:Green1}).  Restricting to $\nu \ge 0$ is sufficient for
our calculation, however, because we can already construct the Green's
functions from these solutions alone; in the formalism of
Refs.~\cite{spheres,universal}, all possible
quantum fluctuations are captured through the Green's function.
Equating Eqs.~(\ref{eqn:Green1}) and
(\ref{eqn:Green2}) and then using (\ref{eqn:plane1}), we obtain the
expansion of the outgoing parabolic solution in plane waves,
\begin{equation}
\psi_\nu^{\hbox{\tiny out}}(\bm{r})
= \frac{e^{ik_z z}}{\sqrt{8\pi}} \int_{-\infty}^{\infty} dk_x
\frac{i}{k_y}
\frac{\left(\tan \frac{\phi}{2}\right)^{\nu}}{\cos \frac{\phi}{2}}
e^{-ik_y y + i k_x x} \,,
\label{eqn:plane2}
\end{equation}
which is valid for $y<0$.

The regular and outgoing waves provide two independent
solutions to the second-order differential equation.  We take a linear
combination of these solutions to obtain the scattering solution
$\Phi_\nu(\bm r)$ outside the parabolic cylinder.  Fixing the
coefficients by imposing Dirichlet boundary conditions at $\mu=\mu_0$,
we obtain
\begin{equation}
\Phi_\nu(\bm r) = 
D_{-\nu-1}(\tilde \mu_0) \psi_\nu^{\hbox{\tiny reg}}(\bm{r})
- i^{\nu} D_\nu(i \tilde \mu_0) \psi_\nu^{\hbox{\tiny out}}(\bm{r}) \,,
\label{DirichletMode}
\end{equation}
while for Neumann boundary conditions we have
\begin{equation}
\Phi_\nu(\bm r) = 
D_{-\nu-1}'(\tilde \mu_0) \psi_\nu^{\hbox{\tiny reg}}(\bm{r})
- i^{\nu+1} D_\nu'(i \tilde \mu_0) \psi_\nu^{\hbox{\tiny out}}(\bm{r}) \,.
\label{NeumannMode}
\end{equation}

These solutions to the Helmholtz equation can be used to compute the 
Casimir forces between a parabolic cylinder and other simple objects,
for example an infinite plate, as depicted in Fig.~\ref{fig:tilt}.
If both objects are perfect mirrors, translational symmetry along the
$z$-axis enables us to decompose the electromagnetic field into two
scalar fields, with Dirichlet and Neumann boundary conditions
respectively.  Each scalar field can then be treated independently,
with the sum of their contributions giving the full electromagnetic
result.  The quantization of each scalar field is achieved by
integrating the exponentiated action over all configurations of the
field~\cite{GK}.  Constraining the fields to obey the
boundary conditions on each surface leads to an alternative description
involving fluctuating ``charges'' $\rho_{\hbox{\tiny plane}}$ and
$\rho_{\hbox{\tiny cylinder}}$ on the surfaces \cite{spheres,universal}.  
Appropriate multipoles of these charges are
\begin{eqnarray}
Q^P(k_x,k_z,\kappa) &=& \int_{\hbox{\tiny plane}} 
\hspace{-0.04\linewidth} 
dx \, dz \, dt \, e^{-i k_x x-ik_z z+\kappa t} 
\rho_{\hbox{\tiny plane}}(x,z,t) , \nonumber \\
Q_\nu^C(k_z,\kappa) &=& \frac{1}{\sqrt{\sqrt{2\pi} \nu!}}
\int_{\hbox{\tiny cylinder}} 
\hspace{-0.04\linewidth} d\lambda \, dz \, dt\,
e^{-ik_z z+\kappa t}   \nonumber\\
&& \hspace{-0.125\linewidth}
\times \psi_\nu^{\hbox{\tiny reg}} (\lambda, \mu_0) 
\rho_{\hbox{\tiny cylinder}}(\lambda, \mu_0,z,t) (\lambda^2 + \mu_0^2) .
\end{eqnarray}

The action can be decomposed in terms of these multipoles as
${\cal S}=\int_0^\infty d\kappa \frac{L dk_z}{2\pi}
\left[{\cal S}_{PP}+{\cal S}_{CC}+{\cal S}_{CP}  +
\hbox{c.c.}\right]$, with
\begin{eqnarray}
{\cal S}_{PP}(\kappa,k_z) &=& 
\frac{-i}{8 \pi} \int_{-\infty}^\infty \frac{dk_x}{k_y}
Q^P(k_x)^\ast ({\cal F}^P_{k_x})^{-1} Q^P(k_x), \nonumber\\
{\cal S}_{CC}(\kappa,k_z) &=& 
- \frac{1}{2} \sum_{\nu=0}^\infty
{Q_{\nu}^C}^\ast ({\cal F}_{\nu}^C)^{-1} Q_{\nu}^C, \\
{\cal S}_{CP}(\kappa,k_z) &=& \cr
&&\hspace{-0.15\linewidth}
\sum_{\nu=0}^\infty \int_{-\infty}^\infty dk_x 
\sqrt{\frac{i}{16 \pi k_y}}
{\cal U}_{\nu k_x}(d,\theta)
{Q_{\nu}^C}^\ast Q^P(k_x) .\nonumber
\end{eqnarray}
Here ${\cal S}_{PP}$ corresponds to the action for the charges on the
plane, with scattering amplitudes ${\cal F}^P_{k_x} = \pm 1$ for
Neumann and Dirichlet modes respectively.  The corresponding action
for charges on the parabolic cylinder ${\cal S}_{CC}$  can be related
to its scattering amplitudes ${\cal F}^C_{\nu}$~\cite{universal}; from
Eqs.~(\ref{DirichletMode}) and~(\ref{NeumannMode}) we obtain
\begin{eqnarray}
{\cal F}_\nu^C  &=&
-i^{\nu}  \frac{D_{\nu}(i\tilde \mu_0)}{D_{-\nu-1}(\tilde \mu_0)}
\hbox{\quad (Dirichlet),} \cr
{\cal F}_\nu^C &=& -i^{\nu+1} 
\frac{D_{\nu}'(i\tilde \mu_0)}{D_{-\nu-1}'(\tilde \mu_0)}
\hbox{\quad (Neumann).}
\label{eq:T}
\end{eqnarray}
The position and orientation of the parabolic cylinder relative to the
plane enter only through the translation matrix ${\cal U}_{\nu
k_x}(d,\theta)$, which appears in the interaction term 
${\cal S}_{CP}$.  From Eq.~(\ref{eqn:plane2}), we obtain
\begin{equation}
{\cal U}_{\nu k_x}(d,\theta)=
\sqrt{\frac{i}{2 k_y \nu! \sqrt{2\pi}}}
\frac{\left(\tan \frac{\phi + \theta}{2}\right)^{\nu}}
{\cos \frac{\phi + \theta}{2}} e^{i k_y d} \,,
\end{equation}
where $\theta$ is the angle of inclination of the parabolic
cylinder and $d$ is the distance from the focus of the parabola to the
plane, as shown in Fig.~\ref{fig:tilt}.

Integrating over these charge fluctuations gives the Casimir energy 
per unit length as
\begin{eqnarray}
\frac{\cal E}{\hbar c L} &=&  \int_0^\infty
\frac {d\kappa}{2 \pi}
\int_{-\infty}^\infty \frac {dk_z}{2 \pi} 
\log \det \left(\mathbbm{1}_{\nu \nu'} - 
\phantom{\int}\right. \\
&& \left.{\cal F}_{\nu}^C
\int_{-\infty}^\infty \hspace{-8pt} d k_x\,
{\cal U}_{\nu k_x}(d,\theta)
{\cal F}_{k_x}^P
{\cal U}_{\nu' k_x} (d,-\theta)
\right) \nonumber .
\end{eqnarray}
Numerical computations are performed by truncating the 
determinant at index $\nu_{\hbox{\tiny max}}$.  For the numbers quoted
below, we have computed for $\nu_{\hbox{\tiny max}}$ up to $200$ and
then extrapolated the result for $\nu_{\hbox{\tiny max}} \to \infty$,
and in the figures we have generally used $\nu_{\hbox{\tiny max}} =
100$.  We note that the integrals over $\kappa$ and $k_z$ can be
expressed as a single integral in polar coordinates, and for
$\theta=0$ the $k_x$ integral is symmetric and the translation matrix
elements vanish for $\nu+\nu'$ odd.  Since the plane we are
considering is a perfect mirror, ${\cal F}_{k_x}^P$ is independent of
$k_x$ and we can further simplify the calculation for $\theta=0$ using
the integral
\begin{equation}
\int_{-\infty}^\infty dk_x \frac{i}{k_y} 
\frac{\left(\tan \frac{\phi}{2}\right)^{2n}}
{\cos^2 \frac{\phi}{2}} e^{2i k_y d}
= 2 \pi {\rm k}_{-2n-1}(2d\sqrt{\kappa^2 + k_z^2}),
\end{equation}
where ${\rm k}_\ell(u)=\frac{e^{-u}}{\Gamma\left(\frac{\ell}{2}+1\right)}
U(-\frac{\ell}{2},0, 2u)$ is the Bateman ${\rm k}$-function \cite{Bateman},
which is zero if $\ell$ is a negative even integer.  Here 
$U(a,b,u)$ is the confluent hypergeometric function of the
second kind.

\begin{figure}[htbp]
\includegraphics[width=0.95\linewidth]{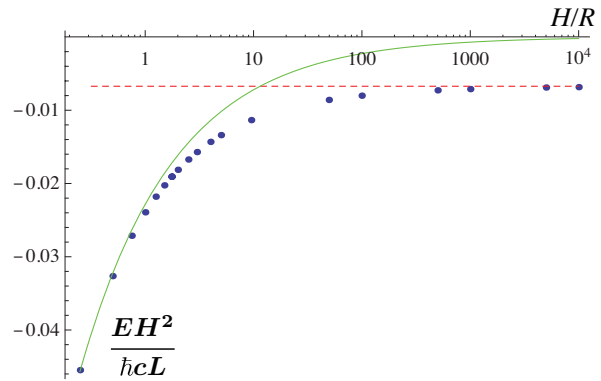}
\caption{The energy per unit length times $H^2$,
$EH^2/(\hbar c L)$, plotted versus $H/R$ for $\theta=0$ and
$R=1$ on a log-linear scale.  The dashed line gives the $R=0$ limit
and the solid curve gives the PFA result.}
\label{fig:EvsH}
\end{figure}

As a first demonstration, we report on the dependence of the energy on
the separation $H = d-R/2$ for $\theta=0$.  At small separations
($H/R\ll 1$) the PFA, given by
\begin{equation}\label{eq:pfa}
\frac{{\cal E}_{\hbox{\tiny pfa}}}{\hbar c L} = 
-\frac{\pi^2}{720} \int_{-\infty}^\infty
\frac{dx}{\left[H + x^2/(2R)\right]^3} = -\frac{\pi^3}{960\sqrt{2}}
\sqrt{\frac{R}{H^5}}\,,
\end{equation}
should be valid.  The numerical results in Fig.~\ref{fig:EvsH} confirm
this expectation with a ratio of actual to PFA energy of $0.9961$ at
$H/R=0.25$ (for $R=1$).  We note that since the main contribution to
PFA is from the proximal parts of the two surfaces, the PFA result in
Eq.~(\ref{eq:pfa}) also applies to a circular cylinder with the same
radius $R$.

A more interesting limit is obtained when $R/H\to 0$, corresponding
to a semi-infinite plate.  Then the PFA result is zero, as are results
based on perturbative approximation for the dilute limit \cite{Milton}.
The scattering amplitudes in Eq.~(\ref{eq:T})
simplify and can be combined together as
${\cal F}_{\nu}^C = -\nu!\sqrt{2/\pi}$,
where even $\nu$ corresponds to Dirichlet and odd $\nu$ corresponds to
Neumann.  Using this result, our expression for the energy 
for $R=0$ and $\theta=0$ simplifies to
\begin{eqnarray}\label{eq:zeroR}
\frac{{\cal E}}{\hbar c L} &=& 
\int_0^\infty \hspace{-1pt}
\frac{q dq}{4\pi} \log \det
\left(\mathbbm{1}_{\nu \nu'} - (-1)^\nu
{\rm k}_{-\nu-\nu'-1}(2 q H)\right)  \cr
&=& \frac{-C_\perp}{H^2},
\end{eqnarray}
where $C_\perp=0.0067415$ is obtained by numerical integration.
This geometry was studied using the world-line method for a scalar field
with {\em Dirichlet} boundary conditions in Ref.~\cite{Gies}.  The
world-line approach requires a large-scale numerical computation,
and it is not known how to extend this method to Neumann boundary
conditions (or any case other than a scalar with Dirichlet boundary
conditions).  In our calculation, the Dirichlet component of the
electromagnetic field makes a contribution $C^D_\perp=0.0060485$ to
our result, in reasonable agreement with the
value of  $C^D_\perp=0.00600(2)$ in Ref.~\cite{Gies}.

\begin{figure}[htbp]
\includegraphics[width=0.95\linewidth]{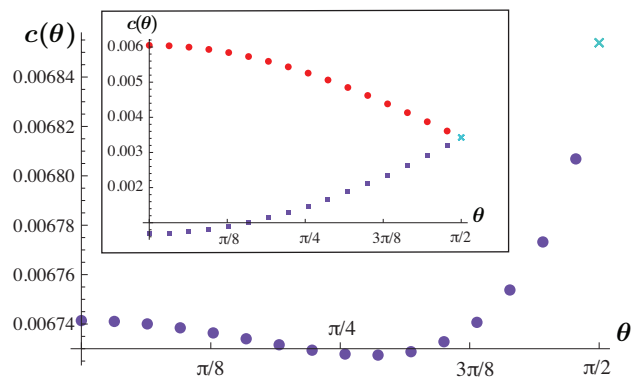}
\caption{The coefficient $c(\theta)$ as a function of angle for $R=0$.
The exact result at $\theta=\pi/2$ is marked with a cross.  Inset:
Dirichlet (circles) and Neumann (squares) contributions to the 
full electromagnetic result.
}
\label{fig:cAmp}
\end{figure}

Reference~\cite{Gies} also considers a tilted semi-infinite plate,
which corresponds to the $R\to0$ limit of our formula for
general $\theta$.  From dimensional analysis, the Casimir energy at
$R=0$ again takes the now $\theta$-dependent form 
\begin{equation}
\frac{{\cal E}}{\hbar c L} = -\frac{C(\theta)}{H^2} \,,
\label{eq:Etheta}
\end{equation}
where $H=d$ for $R=0$.  Following Ref.~\cite{Gies},
we plot $c(\theta)=\cos(\theta)C(\theta)$ in Fig.~\ref{fig:cAmp}.
A particularly interesting limit is $\theta\to \pi/2$,
when the two plates are parallel.  In this case, the leading
contribution to the Casimir energy should be proportional to the area
of the half-plane according to the parallel plate formula,
$E_\parallel /(\hbar cA)= -c_\parallel/H^3$ with
$c_\parallel=\pi^2/720$, plus a subleading correction due to the
edge.  Multiplying by $\cos\theta$ removes the divergence in 
$C(\theta)$ as $\theta\to \pi/2$.  As in Ref.~\cite{Gies}, we assume 
$c(\theta\to \pi/2)=c_\parallel/2+ \left(\theta - \pi/2\right)
c_{\hbox{\tiny edge}}$, although we cannot rule out the possibility of
additional non-analytic forms, such as logarithmic or other
singularities.  With this assumption, we can estimate
the edge correction $c_{\hbox{\tiny edge}} = 0.0009$ from the data in
Fig.~\ref{fig:cAmp}.  From the inset in Fig.~\ref{fig:cAmp}, we
estimate the Dirichlet and Neumann contributions to this result to be
$c_{\hbox{\tiny edge}}^D = -0.0025$ (in agreement with
\cite{Gies} within our error estimates) and 
$c_{\hbox{\tiny edge}}^N = 0.0034$ respectively.  Because higher partial
waves become more important as $\theta \to \pi/2$, reflecting the
divergence in $C(\theta)$ in this limit, we have used larger values of
$\nu_{\hbox{\tiny max}}$ for $\theta$ near $\pi/2$.

It is straightforward to extend these results to nonzero temperature
$T$. We simply replace the integral $\int_0^\infty
\frac{d\kappa}{2\pi}$ by the sum $\frac{T}{\hbar c}
{\sum_{n=0}^\infty} '$ over Matsubara frequencies $\kappa_n = 2 \pi n
T/(\hbar c)$, where the prime indicates that the $n=0$ mode is counted
with a weight of $1/2$ \cite{universal}.  In the limit of infinite
temperature, only the $n=0$ mode contributes and we obtain for $R=0$
the energy ${\cal E}/L = -T C_{T=\infty}/H$, with $C_{T=\infty} =
0.0472$.  The Dirichlet contribution to our result is $C_{T=\infty}^D
= 0.0394$, again in agreement with~\cite{Gies}.

Employing the scattering formalism, we can also calculate the Casimir
energy for the case where another object whose scattering
amplitudes are available, such as an ordinary cylinder or a second
parabolic cylinder, is positioned outside the parabolic cylinder.
Centering the other object at the origin and letting the parabolic
cylinder open downward, with its focus displaced to $y=-d$, we obtain
the necessary translation matrix elements by writing
Eq.~(\ref{eqn:plane2}) for $\bar{\bm{r}}$, where $\bar x = x$, $\bar y
= - y - d$, $\bar z= z$, and then expanding the plane wave on the
right-hand side in the basis appropriate to the other object.  Again
we can allow the parabolic cylinder to tilt by replacing $\phi$ by
$\phi + \theta$ in this expression.  These results can be extended to
multiple objects, as in Ref.~\cite{JamalAlejandro}.  Another
interesting possibility would be to apply the interior Casimir
formalism of Ref.~\cite{interior} an object inside a
parabolic cylinder, potentially extending the results of
Ref.~\cite{Ford,Lombardo}.

The reduction of the parabolic cylinder to a semi-infinite plate
enables us to consider a variety of edge geometries.  A thin metal
disk perpendicular to a nearby metal surface would experience a
Casimir force described by an extension of Eq.~(\ref{eq:zeroR}).
Figure~\ref{fig:EvsH} shows that the PFA breaks down for a
thin plate perpendicular to a plane; the PFA approximation to the
energy vanishes as the thickness goes to zero, while the correct 
result instead has a different power law dependence on the separation.
Based on the full result for perpendicular planes, however, we can
formulate an ``edge PFA'' that yields the energy by integrating
$d\mathcal{E}/dL$ from Eq.~(\ref{eq:zeroR}) along the
edge of the disk.  Letting $r$ be the disk radius, in this
approximation we have $\mathcal{E}_{\hbox{\tiny Epfa}} = -\hbar c C_\perp
\int_{-r}^r (H+ r-\sqrt{r^2-x^2})^{-2} dx \xrightarrow{H/r\to0} -\hbar
c C_\perp \pi \sqrt{r/(2H^3)}$, which is valid if the thickness of
the disk is small compared to its separation from the plane.  (For
comparison, note that the ordinary PFA for a metal sphere of radius
$r$ and a plate is proportional to $r/H^2$.)

A disk may be more experimentally tractable than a plane, since its edge
does not need to be maintained parallel to the plate.  One possibility
would be a metal film, evaporated onto a substrate that either has low
permittivity or can be etched away beneath the edge of the deposited
film. Micromechanical torsion oscillators, which have already been
used for Casimir experiments \cite{Decca07}, seem readily adaptable
for testing Eq.~(\ref{eq:Etheta}).  Because the overall strength of
the Casimir effect is weaker for a disk than for a sphere, observing
Casimir forces in this geometry will require greater sensitivities or
shorter separation distances than the sphere-plane case.  As the
separation gets smaller, however, the dominant contributions arise
from higher-frequency fluctuations, and deviations from the perfect
conductor limit can become important.  While the effects of finite
conductivity could be captured by an extension of our method, the
calculation becomes significantly more difficult in this case because
the matrix of scattering amplitudes is no longer diagonal.

To estimate the range of important fluctuation frequencies, we
consider $R\ll H$ and $\theta=0$.  In this case, the integrand
in Eq.~(\ref{eq:zeroR}) is strongly peaked around $q \approx 0.3/H$.
As a result, by including only values of $q$ up to $2/H$, we still
capture $95\%$ of the full result (and by going up to $3/H$ we include
99\%).  This truncation corresponds to a minimum fluctuation
wavelength $\lambda_{\hbox{\tiny min}} = \pi H$.  For the perfect
conductor approximation to hold, $\lambda_{\hbox{\tiny min}}$ must be
large compared to the metal's plasma wavelength $\lambda_p$, so that
these fluctuations are well described by assuming perfect
reflectivity.  We also need the thickness of the disk to be small
enough compared to $H$ that the deviation from the proximity force
calculation is evident (see Fig.~\ref{fig:EvsH}), but large enough
compared to the metal's skin depth $\delta$ that the perfect conductor
approximation is valid.  For a typical metal film, $\lambda_p \approx
130\ {\rm nm}$ and $\delta \approx 25\ {\rm nm}$ at the relevant
wavelengths.  For a disk of radius $r=100\ \mu{\rm m}$, the present
experimental frontier of $0.1\ {\rm pN}$ sensitivity corresponds to a
separation distance $H\approx 350\ {\rm nm}$, which then falls within
the expected range of validity of our calculation according to these
criteria.  The force could also be enhanced by connecting several
identical but well-separated disks.  In that case, the same force could
be measured at a larger separation distance, where our calculation is
more accurate.

We thank U. Mohideen for helpful discussions.
This work was supported by the National Science Foundation (NSF)
through grants PHY05-55338 and PHY08-55426 (NG), DMR-08-03315 (SJR and MK), 
Defense Advanced Research Projects Agency (DARPA) contract
No. S-000354 (SJR, MK, and TE), by the Deutsche Forschungsgemeinschaft
(DFG) through grant EM70/3 (TE), and by the U. S. Department of Energy
(DOE) under cooperative research agreement \#DF-FC02-94ER40818 (RLJ).


\end{document}